\documentclass[aps,pre,twocolumn,groupedaddress,showpacs,floatfix]{revtex4}
\usepackage{amsmath}
\usepackage{amssymb}
\usepackage{graphicx}

\begin{document}

\title{What are the Best Hierarchical Descriptors for Complex Networks?}

\author{Luciano da Fontoura Costa$^1$, Roberto Fernandes Silva Andrade$^2$}
\affiliation{$^1$Instituto de F\'{\i}sica de S\~ao Carlos.
Universidade de S\~ ao Paulo, S\~{a}o Carlos, SP, PO Box 369,
13560-970, Brazil, luciano@if.sc.usp.br}
\affiliation{$^2$Instituto de F\'isica,
Universidade Federal da Bahia, 40210-340, Salvador, Bahia, Brazil}

\date{\today}

\begin{abstract}

This work reviews several hierarchical measurements of the topology of
complex networks and then applies feature selection concepts and
methods in order to quantify the relative importance of each
measurement with respect to the discrimination between four
representative theoretical network models, namely Erd\"{o}s-R\'enyi,
Barab\'asi-Albert, Watts-Strogatz as well as a geographical type of
network.  The obtained results confirmed that the four models can be
well-separated by using a combination of measurements.  In addition,
the relative contribution of each considered feature for the overall
discrimination of the models was quantified in terms of the respective
weights in the canonical projection into two dimensions, with the
traditional clustering coefficient, hierarchical clustering
coefficient and neighborhood clustering coefficient resulting
particularly effective.  Interestingly, the average shortest path
length and hierarchical node degrees contributed little for the
separation of the four network models.

\end{abstract}

\maketitle

\emph{It is better to know some of the questions than all of the answers.}
(J. Thurber) \vspace{0.5cm}

\section{Introduction}

A relevant analysis of several features of complex systems can be
achieved through the recently developed complex network framework
(e.g.~\cite{Albert_Barab:2002, Dorog_Mendes:2002, Newman:2003,
Boccaletti_etal:2006, Costa_surv:2007}). Due to the large amount of
variables normally involved in such dynamical systems, the set up of a
interaction network, based on functional relationship among its
degrees of freedom, offers a first picture to the actual internal
structure of the system. This process requires the identification of
the pertinent degrees of freedom as nodes, while the edges that
connect them are defined by the mutual influence they are subject
to. The ability of identifying nodes and edges in the appropriate way
is a crucial step in this modeling.

The characterization of the so obtained networks constitute a second
important step in this kind of analysis. In this process, a small
number of features is chosen in order to measure, in an objective way,
pertinent properties of the sets of nodes and edges. The choice of
measurements used in the investigation constitutes the second key
decision during the structural analysis of the networks, as it defines
which information can be obtained (e.g.~\cite{Costa_surv:2007}).
Nowadays, it is consensual that the set of basic measures include the
average number of links per node$\langle k\rangle$, the clustering
coefficient $C$, mean minimal distance among the nodes $\langle
d\rangle$, and the network diameter $D$.  However, such measurements
can not provide a one-to-one characterization of the networks,
i.e. they yield only a degenerate representation from which the
original network can not be recovered.  This is because several
distinct networks may be mapped into the same set of measurement
values. Therefore, new and distinct measures have been proposed in
order to capture new aspects not covered by the set of four parameters
listed above.

One particularly important aspect regards the characterization of
individual nodes in the network, as this allows the identification of
particularly distinct nodes such as hubs (e.g.~\cite{Costa_out:2006}.
While both the degree and clustering coefficient are defined for each
individual node, they provide but a limited characterization of the
connectivity around those nodes, with several nodes resulting with
identical pairs of degree/clustering coefficient values even when they
are placed at completely different contexts in the network.  One
interesting means to obtain a richer (i.e. less degenerate) set of
measurements for each node is to consider subsequent hierarchical
neighborhoods (e.g.~\cite{Faloutsos_etal:1999, Newman:2001,
Cohen_etal:2006, Costa_hier:2004, Costa_Rocha:2005, Costa_Silva:2006,
Costa_Sporns:2006, Andrade:2006}) around each node, in addition to the
immediate neighbors considered in the traditional degree and
clustering coefficient.

Although new measurements provide extra information, it is important
to understand how they are related to those in the basic set and among
themselves. If one defines a measure space whose axes are spanned by
the distinct parameters, one important issue regards the distribution
of observations along each axis. A full answer to such task should
include also an analytical relationship between the co-linear
(correlated) measures. Another related issue refers to deciding, given
a set of distinct measures, which of them are most effective in
identifying and discriminating between distinct kinds of networks. The
purpose of this work is to address these questions, by working with a
set of hierarchical measures and using sound concepts and methods of
multivariate statistics (e.g.~\cite{Costa_surv:2007, Duda_Hart:2001,
McLachlan:2004}). We probe a large number of networks generated
according to four representative theoretical models, Erd\"{o}s-R\'enyi
(ER), Barab\'asi-Albert (BA), Watts-Strogatz (WS) as well as a
geographical type of network (GG), which can be put in connection to
distinct complex network paradigmatic types, displaying feature that
are associated to the random, scale-free and small-world behaviors. In
this way, it becomes possible to quantify the relative importance of
each measurement, with respect to the discrimination between the
considered distinct network models.  Although illustrated for these
specific four types of networks, the reported methodology is
completely general and can be applied virtually to any problem
involving the choice of measurements given specific types of
theoretical or real-world networks.

The hierarchical measurements we take into account have been discussed
in a series of previous investigations (e.g.~\cite{Costa_hier:2004,
Costa_Rocha:2005, Costa_Silva:2006, Andrade:2006}), in which the
authors have inquired how a node sees not only its immediate
neighborhood, but also successive neighborhoods up to a maximal
distance $D$ from the reference node. The concepts of hierarchies and
higher order neighborhoods, that have been independently introduced
(e.g.~\cite{Faloutsos_etal:1999, Newman:2001, Cohen_etal:2006,
Costa_hier:2004, Costa_Rocha:2005, Costa_Silva:2006,
Costa_Sporns:2006, Andrade:2006}), aim at providing a description of
the relationship among given sets of nodes which are not necessarily
linked by immediate edges, but for which the minimal distance along
the network is bound to a value $1\leq\ell\leq D$. $\ell$ dependent
clustering coefficients and node degrees have been investigated and
compared for several sets of networks. In a second line of
investigation, a recent contribution raises the issue of the
interdependence among distinct measures, while reviewing the most
relevant measures that have been introduced so
far~\cite{Costa_surv:2007}. The results reported herein, heavily based
on the ideas developed in the quoted references, are aimed at
quantifying the role of the several hierarchical measurements while
discriminating between the four considered theoretical network models.
In order to quantify the influence of each measurement on the
separation between the four classes of networks, we apply sound and
objective concepts from multivariate statistics, namely
standardization and canonical projections
(e.g.~\cite{Costa_surv:2007,McLachlan:2004}).

This work is organized as follows: In Section II, we present the basic
notions of complex networks and of the theoretical models used in our
investigation. In Section III, we discuss the hierarchical
measurements that will be taken into account for the selection
method. These methods are presented and discussed in Section
IV. Results from our analyzes are discussed in Section V, while
Section VI closes the work with the concluding remarks.

\section{Basic Concepts}

This section introduces the main concepts used in our analysis,
including network representation as well as the four theoretical
models.

\subsection{Complex Networks Basic Concepts}

A non-weighted complex network $\Gamma$ with $N$ nodes and $E$ edges
can be fully specified in terms of its \emph{adjacency matrix} $K$, so
that $K(i,j)=1$ indicates the existence of an edge extending from node
$j$ to node $i$.  All networks considered in this work are undirected,
which implies $K$ to be symmetric.  They are also devoid of multiple
or self-connections.  Networks whose nodes have well-defined spatial
positions within an embedding space are called \emph{geographical
networks}.

The \emph{degree} $k_i$ of a node $i$ is defined as the number of edges
connected to it. In case a node $j$ can be reached from a node $i$, we can
say that there is a \emph{path} between these two nodes. Two nodes can be
connected through more than one distinct path. The \emph{shortest path}
$d_{i,j}$ between two nodes $i$ and $j$ corresponds to the path with the
smallest number of edges connecting those nodes. The \emph{immediate
neighborhood} of a node $i$ is the set of nodes which are directly
connected to $i$, i.e. the nodes $j$ for which $d_{i,j}=1$. The average
shortest path $\langle d_i \rangle$ to a node $i$ is the mean value of
$d_{i,j}$ over all nodes $i\neq j$, while the network average shortest
path $\langle d \rangle$ is obtained by taking the mean value of $\langle
d_i \rangle$ over the whole set of network nodes.

The \emph{clustering coefficient} $C_i$ of node $i$ can be calculated as
the ratio between the number of edges among the immediate neighbors of $i$
and the maximum possible number of edges between those nodes. Although
measurements such as the node degree and clustering coefficient apply to
individual nodes, it is common to take their average along the network,
yielding the \emph{average node degree} $\left< k \right>$ and the
\emph{average clustering coefficient} $C$.

\subsection{Complex Networks Models}

Several theoretical models of complex networks have been proposed
(e.g.~\cite{Albert_Barab:2002, Dorog_Mendes:2002, Newman:2003,
Boccaletti_etal:2006, Costa_surv:2007}).  As announced in Section I,
the current work considers four of such models (ER, BA, WS and GG),
the most important features of which we briefly describe below. All
networks used for the comparison of the hierarchical measurements in
this work have the same number of nodes $N$ and average node degrees
$\left< k \right>$ as similar as possible.

The ER model (e.g.~\cite{Albert_Barab:2002}) is characterized by having
constant probability $\rho$ of connection between any possible pair of
nodes. Its average degree is given as $\left< k \right> = 2E/N = 2(N-1)
\rho$.  The BA model can be obtained by starting with randomly
interconnected $m0$ nodes.  At each subsequent step, a new node is
connected to $m$ nodes in the current network such that each
connection is preferential to the degree of the previous nodes.  The
average degree of a BA model is given as $2m$.  Therefore, in order to
have ER and BA networks with the same node degree, we need to enforce
that $m = (N-1)\rho$.  WS networks can be produced by starting with
the $N$ nodes distributed along a ring and connecting each node to its
$\left< k \right>/2$ clockwise neighbors and to the same number of
counterclockwise neighbors, with $\left< k \right>$ being an even
number. Then, a small percentage of edges are randomly rewired.
Finally, the geographical model considered in this work is obtained by
considering a Poisson spatial distribution of points with density
$\gamma$ in a two-dimensional embedding space with uniform connecting
all pair of nodes which are at Euclidean distance smaller than
$\sqrt{\left< k
\right>/(\gamma \pi)}$.

\section{Hierarchical Measurements}

Two of the most ubiquitously accepted network measures, namely the
average number of links per node$\langle k\rangle$ and the clustering
coefficient $C$, reflect the immediate landscape of the nodes, as they
just consider, respectively, the number of neighbors each node is
connected to by a direct edge, and how the neighbors of a node are
connected among themselves. The hierarchical measurements introduced
in \cite{Costa_hier:2004, Costa_Rocha:2005, Costa_Silva:2006,
Costa_Sporns:2006, Andrade:2006} first require the identification of
the sets named the \emph{hierarchical shells} $H_i(\ell)$ or,
alternatively, the neighborhoods $N_i(\ell)$, of order $\ell$ of a
node $i$ as the nodes that lie at a minimal distance $\ell$ along the
edges of the network of a given node $i$. For the sake of uniqueness,
from now on we call these sets as $H_i(\ell)$. The hierarchical
measurements result from the extension of the two basic concepts to
the sets $H_i(\ell)$.

For the feature selection analysis we consider, respectively, two and
three distinct types of node degrees and clustering coefficients,
which are so defined. The \emph{average degree}
\begin{equation}\label{eqa1}
\langle k (\ell)\rangle=\sum_{i=1}^{n} k_i(\ell),
\end{equation}
where $k_i(\ell)$ counts the number of neighbors which are at a minimal
distance $\ell$ of node $i$, indicates how the higher order neighborhoods
of each node are populated. The \emph{average hierarchical degree}
\begin{equation}\label{eqa2}
\langle k^H(\ell)\rangle=\sum_{i=1}^{n} k_{i}^{H}(\ell),
\end{equation}
has a different meaning, as $k_{i}^{H}(\ell)$ counts the number of links
between elements of the two sets $H_i(\ell)$ and $H_i(\ell+1)$. It
expresses how deep connected are the nodes that lie in two successive
hierarchical shells, namely $\ell$ and  $\ell+1$, of node $i$. Observe
that we have $\langle k(\ell=1)\rangle=\langle k^H(\ell=0)\rangle=\langle
k\rangle$, if we consider that the $0$-th order neighborhood of a node is
the node itself.

The three distinct $\ell$ dependent clustering coefficients coincide with
the usual $C$ when $\ell=1$. The \emph{hierarchical clustering
coefficient} $C^{H}(\ell)$ counts how many of the
$k_i(\ell)(k_i(\ell)-1)/2$ pairs of nodes formed the elements of the set
$H_i(\ell)$ are directly linked by one edge. On the other hand, the
\emph{neighborhood clustering coefficient} $C^{N}(\ell)$ takes into
account those pairs of the same set that are neighbors of order
$\ell$.  The original clustering coefficient $C$ is a direct measure
of the presence of nearby triangles in a network, and it indirectly
hints to the presence of connected structures as cliques. The higher
order $C^H(\ell)$ and $C^N(\ell)$ give information on the how the
nodes on more distinct hierarchical shells are related among
themselves.

Finally, the \emph{hierarchical clustering coefficient by balls}
$C^{B}(\ell)$, which was also previously introduced, constitutes the third
measurement that takes into account the hierarchical structure of
neighbors of a node. This is a cumulative measure in the sense that,
instead of considering the nodes in a single set $H_i(\ell)$, it considers
all nodes in the set $\mathfrak{H}_i(\ell)=\bigcup_{i=1}^{\ell}H_i(\ell)$.

To evaluate all the above hierarchical measures we profited from the
formalism introduced in \cite{Andrade:2006}, which amounts to first
identifying all the higher order neighborhoods of the networks and
storing the information in a single matrix
\begin{equation}\label{eqa3}
\mathbf{\widehat{M}}=\sum_{\ell=0}^{D}\ell M(\ell).
\end{equation}
All distinct hierarchical measures can be easily defined in terms of the
elements of $\mathbf{\widehat{M}}$.

\section{Feature Selection Methods}

Given $L$ classes of networks (in the case of the current article the
four theoretical models ER, BA, WS and GG) and $Q$ respective
measurements of their topology, an important question is: which subset
of measurements is more effective for discriminating between such
classes?  Such a problem provides a good example of \emph{feature
selection}.

Two main approaches have been considered for feature selection:
\emph{filter} and \emph{wrapper}.  The difference between these two
families of methods is that the latter evaluates the features by
considering the results obtained after feeding them into a classifier,
while the former methods investigate the intrinsic relationship between
the measurements between and/or within the classes
(e.g.~\cite{Han_Kamber:2001, Hand_etal:2001, McLachlan:2004})).  For
instance, the canonical projection method used in this work provides
an example of a filter approach to feature selection.

It should be observed that none of the feature approaches are
absolutely optimal. While wrapper methods will select features which
are most effective for given classifiers, filter approaches will
depend on the definition of some optimality criterion.  For instance,
the canonical projection method adopted in this work quantifies the
separation between the classes by maximizing the distance between the
classes and minimizing the dispersion inside each class (see
Section~\ref{sec:canon}).  Because our interest in the current work is
to characterize the discrimination power of the several hierarchical
measurements, we limit our attention to filter feature selection
methods.

The following subsections present the basic concepts from multivariate
statistics as well as the principal component analysis and the
canonical projection methodologies.

\subsection{Basic Concepts in Multivariate Statistics}
\label{sec:multivar}

Let each of the $Q$ objects of interest (e.g. networks) be
characterized in terms of $R$ measurements $x(i)$, $i = 1, 2, \ldots,
R$. It is convenient to organize the set of measurements obtained for
each object $p = 1, 2, \ldots, Q$ into the respective \emph{feature
vector}

\begin{equation}
  \vec{v}_p = \left[ x_p(1), x_p(2), \ldots, x_p(R)  \right]^T.
\end{equation}

The \emph{mean feature vector} $\vec{\mu}$ can be calculated as

\begin{equation}
  \mu(i)  = \frac{1}{Q} \sum_{p=1}^{Q} x_p(i).
\end{equation}

The elements $C(i,j)$ of the \emph{covariance matrix} $C$ of the
measurements of the objects can be estimated as

\begin{equation}
  C(i,j) = \frac{1}{Q-1} \sum_{p=1}^{Q} (x_p(i) - \mu(i)) (x_p(j) - \mu(j))
\end{equation}

The \emph{standardized} feature vector can be obtained as

\begin{equation}
  \vec{s}_p = \left[\frac{x_p(1) - \mu(1)}{\sigma(1)}, \frac{x_p(2) -
  \mu(2)}{\sigma(2)}, \ldots, \frac{x_p(R) - \mu(R)}{\sigma(R)}  \right]^T,
\end{equation}
where $\sigma({i})$ is the standard deviation of measurement $x(i)$. Note
that each normalized measurement $s(i)$ has zero mean and unity standard
deviation.

The \emph{Pearson Correlation Coefficient} between two measurements
$x(i)$ and $x(j)$ can be given by the covariance between the
standardized measurements $s(i)$ and $s(j)$.

\subsection{Principal Component Analysis --- PCA}

The multivariate statistical method known as principal component
analysis (PCA) allows dimensionality reduction while maximizing the
data variance along the first projected axes
(e.g.~\cite{Duda_Hart:2001, Costa_surv:2007, Costa_Sporns:2006}).
Because the class of each point is not taken into account in this
method, it corresponds neither to filter nor wrapper feature
selection.  This method is considered in this work for two reasons.
First, it can be used to obtain preliminary visualizations of the
distribution of points and classes.  Second, it provides an
introduction and a comparison standard to the more sophisticated
canonical projections methodology, to which it is related.

Given the set of $Q$ objects, characterized by $R$ measurements, it is
possible to project such measurements into a reduced space with $W <
Q$ dimensions.  In order to do so, the covariance matrix $C$ of the
measurements is estimated as described in Section~\ref{sec:multivar}
and its eigenvalues and respective eigenvectors are calculated.  The
eigenvectors corresponding to the $W$ largest eigenvalues (in
decreasing order of absolute values) are organized into a matrix $A$
such that each line corresponds to an eigenvector.  The matrix $A$
defines the statistical linear transformation of the original set of
data that maximizes variances along the first new axes. Provided $W$
is equal to 2 or 3, the so-transformed data can now be visualized as a
2D or 3D distribution of points.  The original classes of each point
can be visualized with different marks.

\subsection{Canonical Projections} \label{sec:canon}

The method of canonical projections, also called canonical analysis or
canonical variables (e.g.\cite{Duda_Hart:2001, McLachlan:2004}), also
performs a projection of the original measurement space, but now
considering explicitly the original classes of each object.  The
projection is performed not in order to maximize the variances along
the first new axes, but so as to obtain maximum separation of the
classes, quantified by an optimality index $\xi$ reflecting the
distribution of the data both inside and among classes.  More
specifically, $\xi$ will favor well-separated classes, with small
dispersions of the respective objects.  The inter- and intra-class
dispersion matrix, respectively $D_e$ and $D_a$, can be calculated as
described in~\cite{Costa_book:2001, Duda_Hart:2001, Costa_surv:2007}.
The eigenstructure of the matrix $(D_a)^{-1}D_e$ provide the basis for
the sought linear transformation (as in the PCA, the eigenvectors
associated to the largest absolute eigenvalues are stacked as lines in
the transformation matrix) maximizing the separation between the
classes.  For instance, in the case of canonical projections into
two-dimensional spaces, the eigenvectors $v1$ and $v2$ associated to
the largest and second largest absolute eigenvalues are used to define
the projection linear transformation.  Similarly to the PCA, the
contribution of each original measurement to the projection can be
quantified in terms of the absolute value of the weights defined by
the respective eigenvector transformation matrix. Therefore, the
measurements yielding the largest absolute weights for the first axes
can be understood as those which are more important for the separation
between the classes.  In this work we define the \emph{importance} of
each measurement $i$ as the sum of the absolute values of the
respective weights in $v1$ and $v2$, i.e.

\begin{equation}
  I(i) = | v1(i) | + | v2(i) |
\end{equation}

In order to avoid intrinsic biases implied by the relative amplitude
of each measurement, it is interesting to perform the canonical
projections on standardized versions of the measurements.

\section{Results and Discussion}

In order to investigate, in a comparative fashion, the relative
contributions of each measurement for the characterization and
discrimination between the four considered complex network models, 30
realizations of each model, all with mean degree equal to 6 and sizes
$N$ of 100, 200 and 300 nodes, were first obtained.  Two types of WS
networks were obtained, considering $0.1 N$ and $0.1 E$ connection
rewirings, where $E$ is the overall number of connections. These two
types of WS networks are henceforth abbreviated as WS-R and WS-S.
Observe that the former type corresponds to almost regular networks
(i.e. similar node degrees throughout), while the latter type presents
the small world property. All considered models had their average
traditional and hierarchical measurements (for $\ell=1,2,$ and $3$)
calculated and used as feature vectors. Table~\ref{tab:abbrs} lists
the considered measurements as well as their respective symbols and
abbreviations.

\begin{table}[htb]
\centering
\begin{tabular}{||l|c|c||} \hline
 Measurement & Symbol & Abbrev- \\
  &  & iation \\ \hline
 Hierarchical clustering coefficient by balls & $C^B(\ell)$ &  $cb$ \\
 Hierarchical clustering coefficient &  $C^H(\ell)$ & $cl$ \\
 Neighborhood clustering coefficient &  $C^N(\ell)$ & $cn$ \\
 Average number of nodes & $\langle k(\ell) \rangle$ & $n$  \\
 Average shortest path  &  $\langle d \rangle$ & $sp$ \\
 Average hierarchical degree & $\langle k^H(\ell) \rangle$ & $hd$ \\  \hline
\end{tabular}
\caption{The symbols and abbreviations of the considered hierarchical
measurements defined in Sections II and III. The respective
hierarchical level $(\ell)$ is henceforth represented in front of each
abbreviation, e.g. the hierarchical node degree at level $\ell=3$ is
abbreviated as $hd3$}.\label{tab:abbrs}
\end{table}

\begin{figure}
  \centerline{\includegraphics[width=1\linewidth]{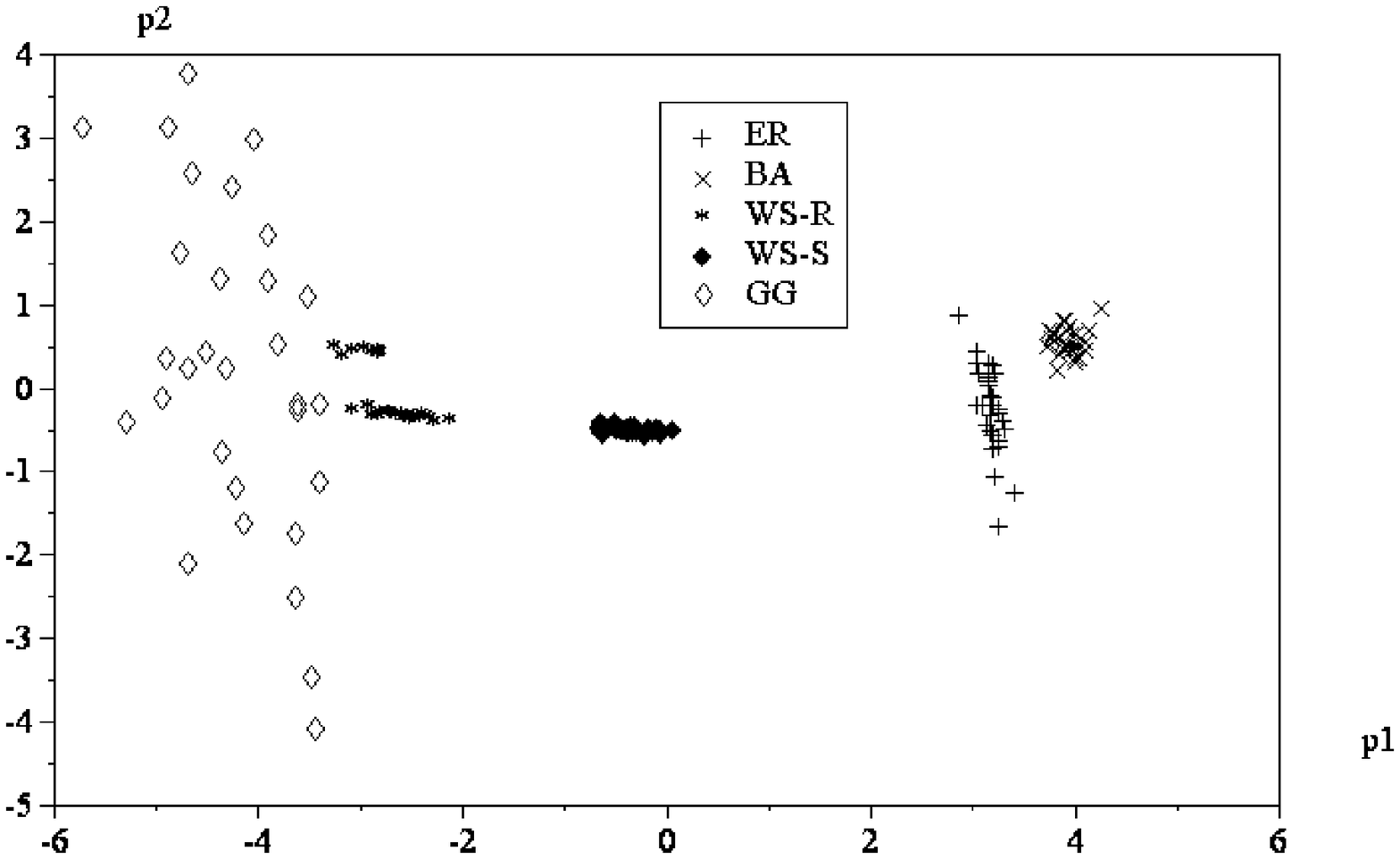}}
  (a) \\
  \centerline{\includegraphics[width=1\linewidth]{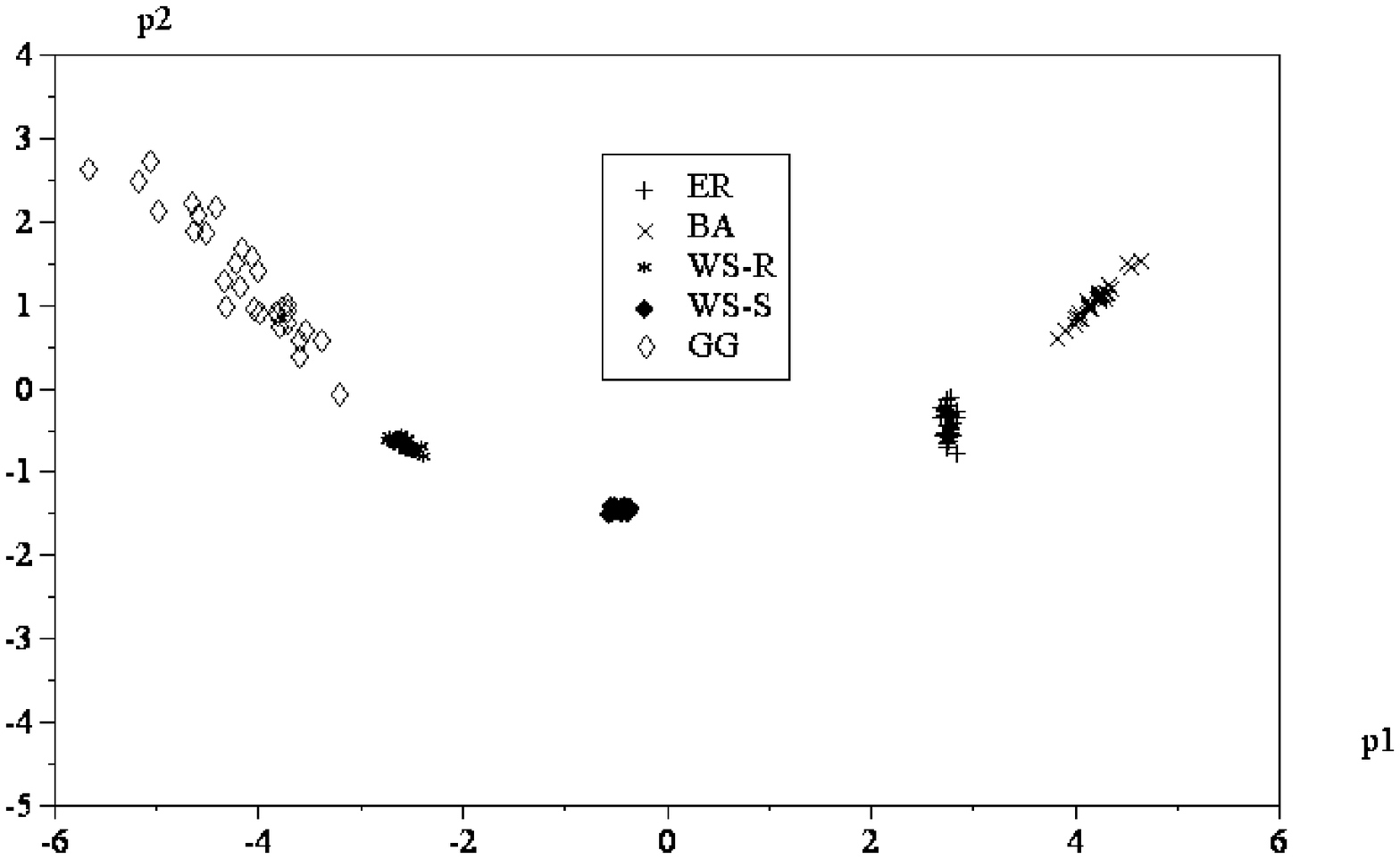}}
  (b) \\
  \caption{The distribution, in the PCA projected phase space, of
  the four theoretical complex network models obtained for
  networks with 100 (a) and 300 (b) nodes. The axes $p1$ and
  $p2$ correspond to the two main projection orientations
  as provided by the eigenvectors associated to the largest
  absolute eigenvalues of the covariance matrix.}
  \label{fig:pca}
\end{figure}

Figure~\ref{fig:pca} shows the two-dimensional phase spaces obtained by
PCA projection of the original 13-dimensional phase spaces for $N=100$ and
$300$.  Each point in this phase space corresponds to a specific network
realization.  The ER and BA clusters resulted near one another, which was
also obtained for the WS/GG pair of clusters. The GG networks resulted in
the most dispersed cluster.

\begin{figure}
  \centerline{\includegraphics[width=1\linewidth]{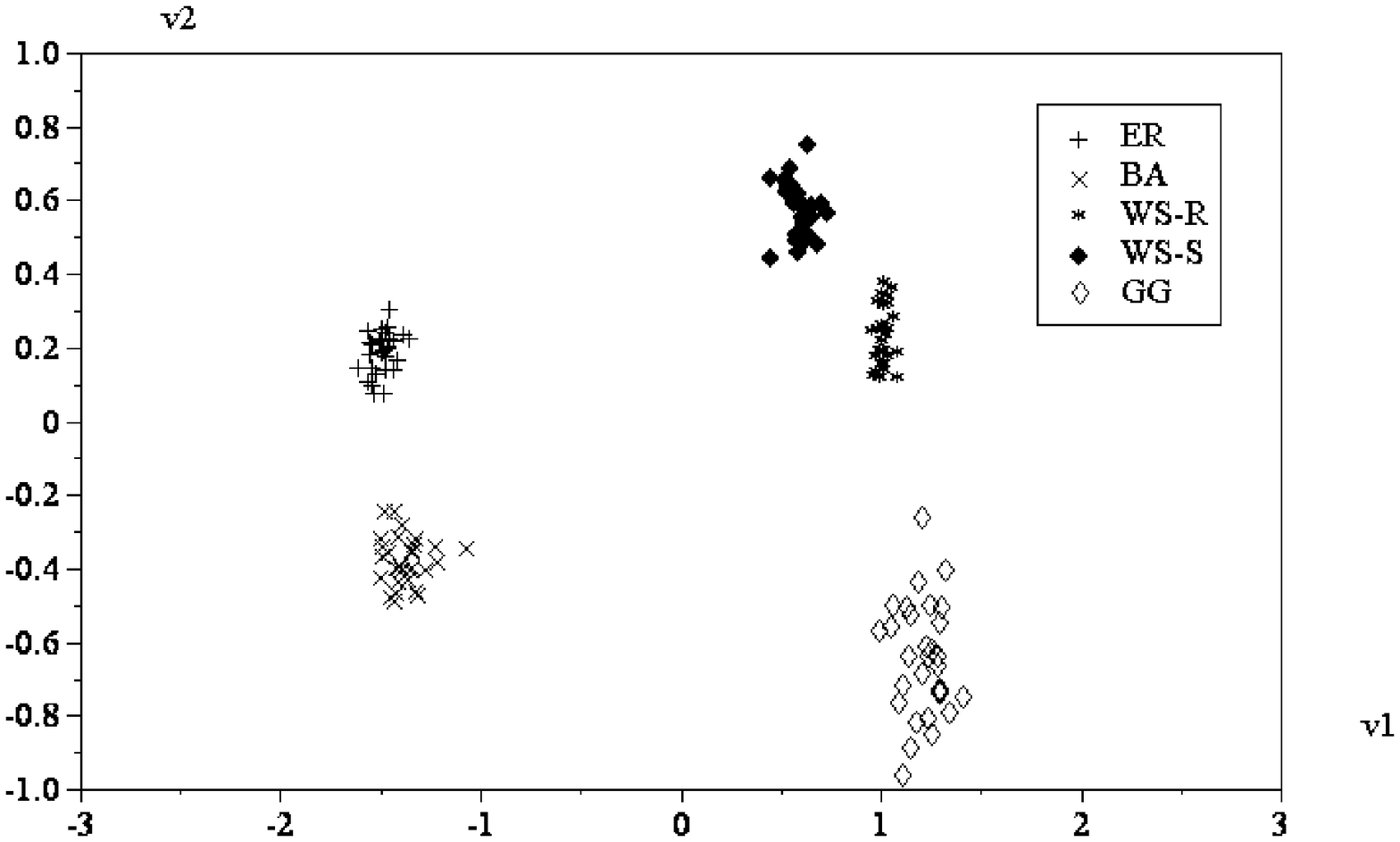}}
  (a) \\
  \centerline{\includegraphics[width=1\linewidth]{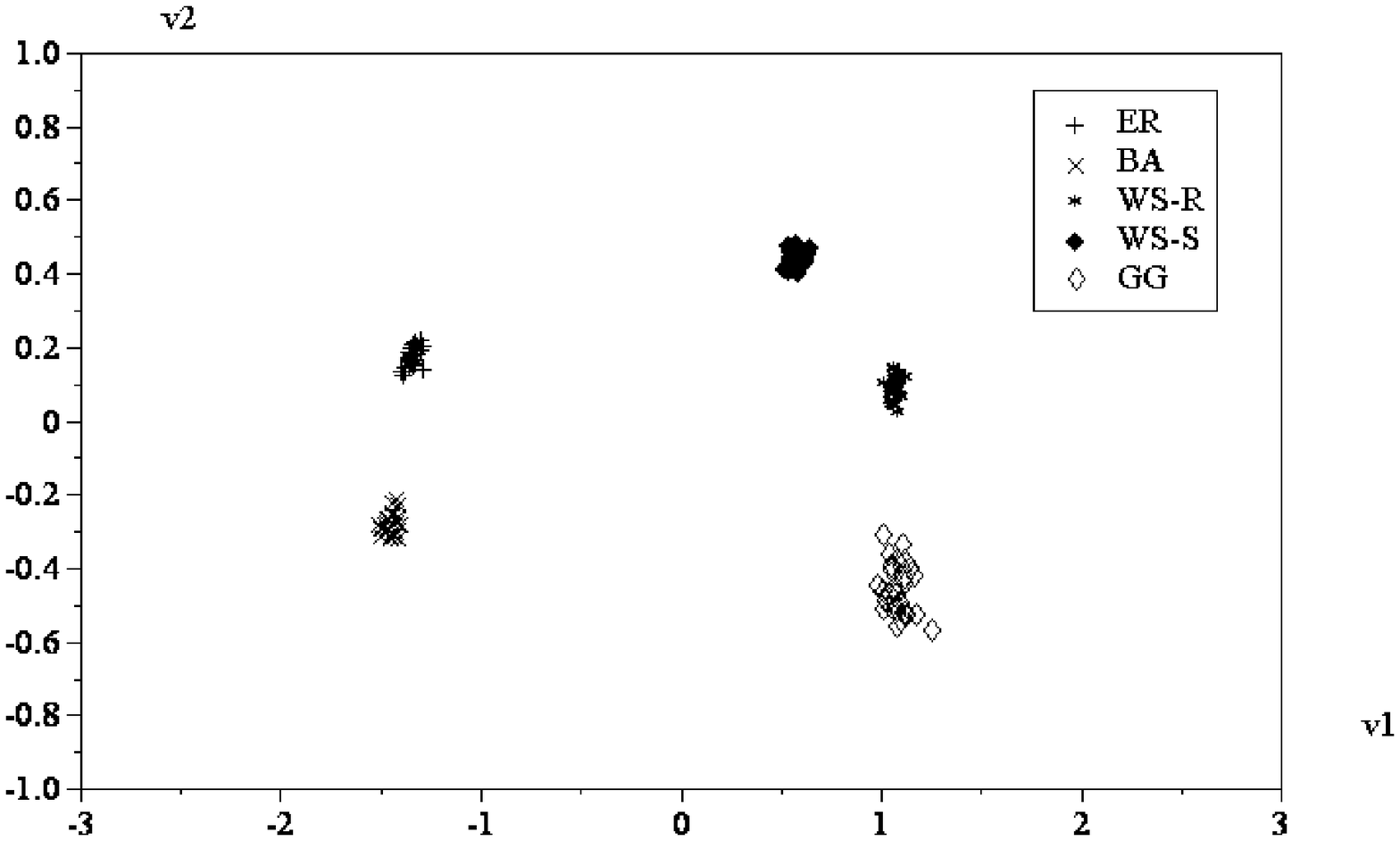}}
  (b) \\
  \caption{The distribution, in the projected phase space, of
  the four theoretical complex network models obtained for
  networks with 100 (a) and 300 (b) nodes. The axes $v1$ and
  $v2$ correspond to the two main projection orientations
  as provided by the eigenvectors associated to the largest
  absolute eigenvalues.}
  \label{fig:scatts}
\end{figure}

The phase spaces obtained while considering all 13 measurements were
also projected into two dimensions by using the canonical methodology
described in Section~\ref{sec:canon}.  Figure~\ref{fig:scatts} shows
the projected phase spaces obtained for networks with size of $N=100$
(a) and 300 (b) nodes, respectively.  It is clear that the separation
between the four networks modes is much better than that obtained by
using PCA (Fig.~\ref{fig:pca}).  It is also clear from the two
dimensional spaces in Fig.~\ref{fig:scatts} that the four models could
be very well separated as a consequence of using such a comprehensive
set of features. Interestingly, the ER/BA and WS/GG models again
tended to cluster together.  Observe that the dispersion of the points
for all distinct classes decreased for larger $N$. The relative
separation between the ER and BA models along the $v_2$ direction also
decreased for this case.  Also, the WS-R and WS-S families of networks
resulted near one another.

\pagebreak

\begin{figure}
  \centerline{\includegraphics[width=1\linewidth]{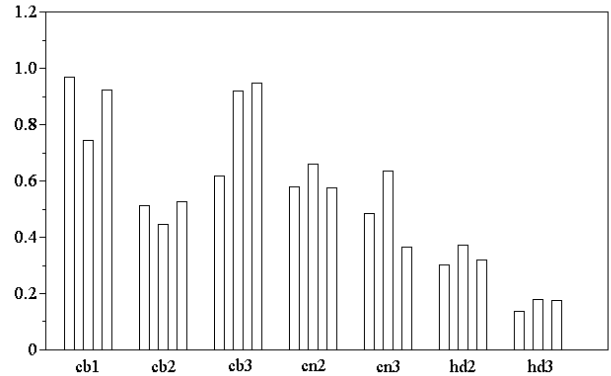}}
  \caption{The importance of the most relevant measurements
           for each of the considered three network sizes, i.e.
           100, 200 and 300 nodes.}
  \label{fig:comp_bars}
\end{figure}

The seven most important measurements considering all the three
network sizes, in decreasing order, were: $cb1=ch1=cn1=C$, $cb3$,
$cn2$, $cn3$, $cb2$, $cb3$ and $hd2$.  Figure~\ref{fig:comp_bars}
shows the importance of each of these measurements obtained for each
of the network sizes (i.e.  100, 200 and 300). It is noteworthy the
absence of two of the most used measurements, ($\langle k \rangle $
and $\langle d \rangle$, as well as of the higher order hierarchical
node degrees, for the purpose of identifying the distinct network
classes.  The fact that all networks considered in this work had
nearly the same average degree explains why $\langle k
\rangle$ had little contribution for the discrimination. However, the fact
that the distribution of average hierarchical degrees have been found to
vary between different network models~\cite{Costa_Silva:2006} should, at
least in principle, imply better discrimination potential for those
measurements.  The relatively minor contribution provided by the average
shortest path is also surprising.

The traditional clustering coefficient, $C$ resulted the most
important measurement in all cases, but the relevance of considering
higher hierarchies for the characterization of the considered networks
was corroborated by relatively high importance obtained for the other
measurements. Interestingly, the importance of the clustering
coefficient by balls for hierarchy 3 tended to increase with $N$,
while the neighborhood clustering coefficient for hierarchy 3
decreased with that parameter. The former effect is a consequence of
the fact that the hierarchical degree becomes more relevant in larger
networks, because such networks have larger diameter and therefore
allow more elaborated and unfolded neighborhoods.

\begin{figure*}
  \centerline{\includegraphics[width=0.95\linewidth]{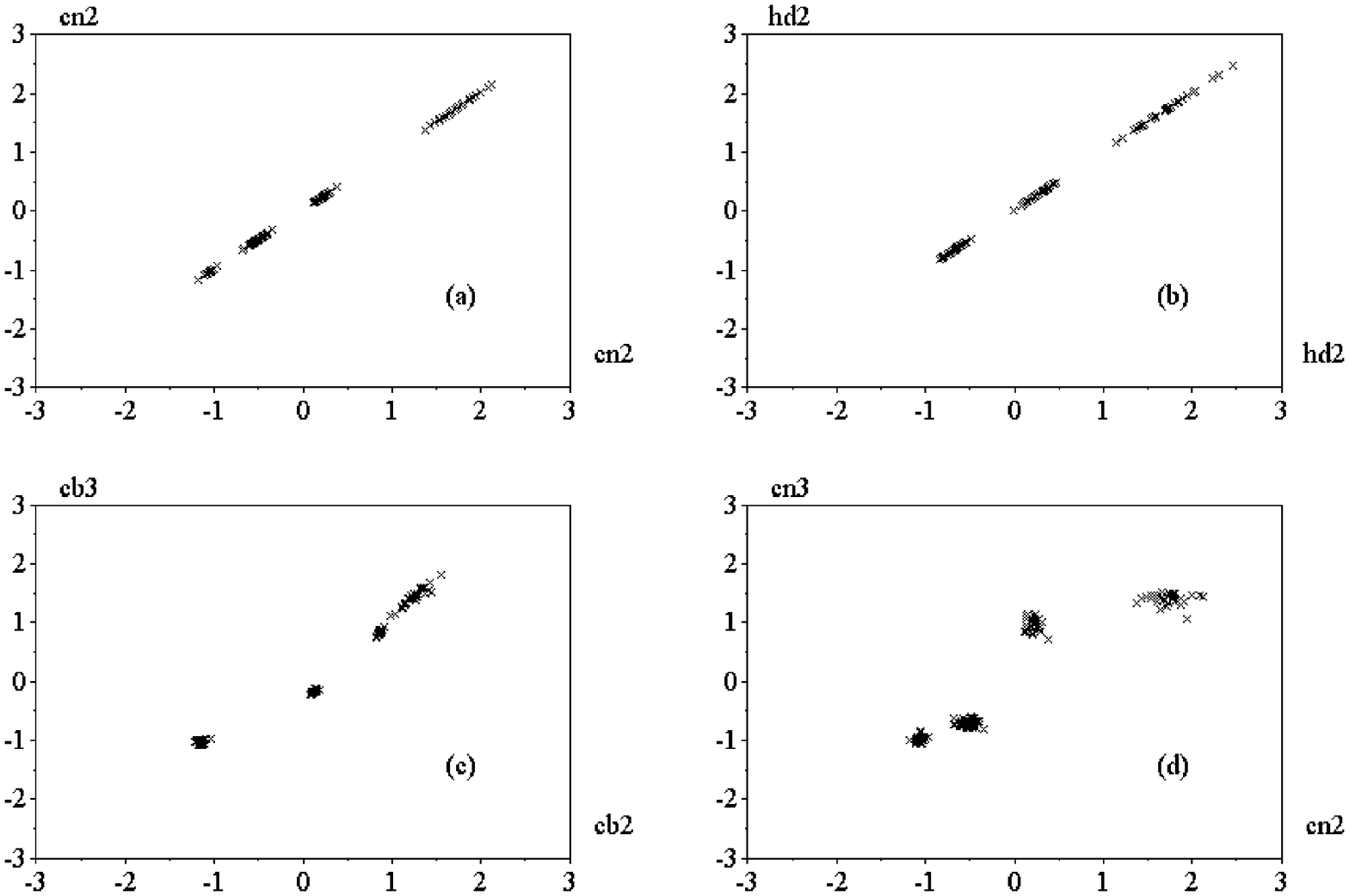}}
  \centerline{\includegraphics[width=0.95\linewidth]{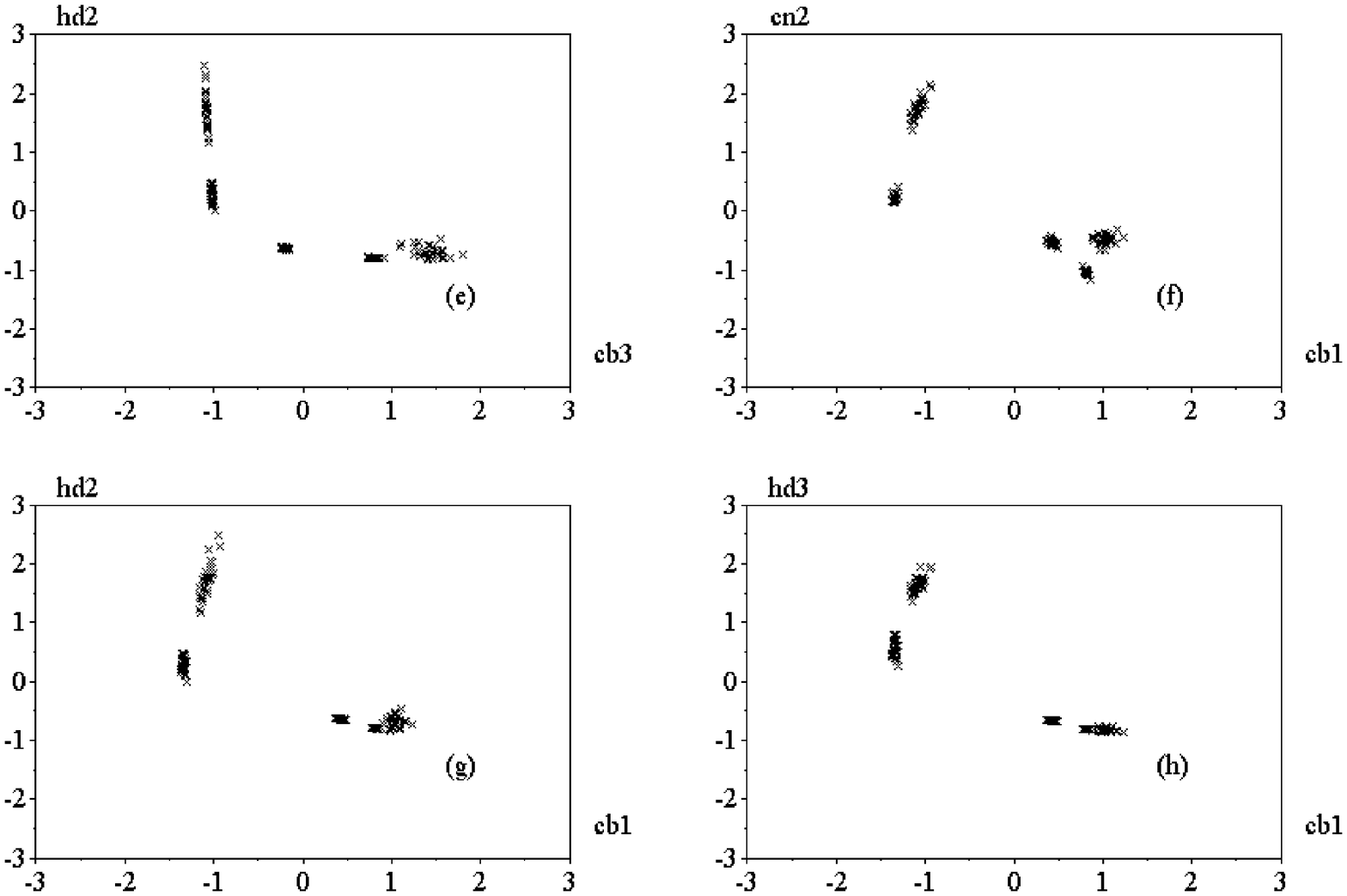}}
  \caption{Scatterplots respective to pairwise combinations of the
  adopted measurements for $N=300$.  Measurements $cn2$ and $hd2$
  provided the best discrimination between the network models when taken
  individually (a,b).  Measurements taken at successive hierarchies tended to
  be moderately correlated (c), while distinct types of measurements
  presented very little correlation (d-h).}  \label{fig:corrs}
\end{figure*}

Figure~\ref{fig:corrs} illustrates several scatterplots obtained
considering pairs of the adopted measurements.
Figures~\ref{fig:corrs}(a,b), corresponding to $C^N(2) \times C^N(2)$ and
$k^H(2) \times k^H(2)$, show that these two measurements provided good
discrimination between the four network models. For other measurements
(not shown), we can observe the existence of fewer distinct disjoint
regions.  The other scatterplots in Figure~\ref{fig:corrs} correspond to
pairwise associations between distinct measurements. For instance, the
panel $C^B(2) \times C^B(3)$ (Fig.~\ref{fig:corrs}c) has most of the
points aligned along the diagonal, indicating that these measures provide
almost the same kind of information (statistically, the two measurements
are said to be correlated). A similar tendency was observed for the same
hierarchical measurements with distinct values of $\ell$, as in the panels
for $C^N(2) \times C^N(3)$ (Fig.~\ref{fig:corrs}d) and $\langle
k^H(2)\rangle \times \langle k^H(3)\rangle$ (Fig.~\ref{fig:corrs}e). Note,
however, that the points in these cases are aligned in a less clear way in
comparison to $C^B(2) \times C^B(3)$ (Fig.~\ref{fig:corrs}c). The panels
which combine hierarchical measures of distinct classes $(C^H, \langle k^H
\rangle, C^N)$ have the points away from the diagonal (e.g.
Fig.~\ref{fig:corrs}e-h). This is a indication that these measurements are
uncorrelated, tending to provide non-redundant information and,
consequently, enhanced discrimination power.  It is however important to
stress that the overall discrimination can not be fully predicted simply
from pairwise relationships between the measurements, such as those
illustrated above. Observe also that none of the two-dimensional
scatterplots in Figure~\ref{fig:corrs} provide separation between the four
models as good as that shown in Figure~\ref{fig:scatts}b.  That is because
the latter scatterplot was obtained from the much higher dimensional phase
space by projecting into the plane allowing the best separation between
the four models. Such a result clearly corroborates the increased
separability allowed by the consideration of a comprehensive set of
distinct hierarchical measurements.

Finally, we proceeded with a further test to investigate the ability
of the used feature section methods to uncover network
specificities, by adding one extra network to two of the previous
groups of 120 specimens. We considered the Apollonian network (AN)
\cite{Andrade:2005, Doye:2005}, a geometrical assembly of nodes and
links, which is defined on the basis of the classical problem of
finding the optimal covering of a plane by circles. As they are
defined in a recursive way, we considered two successively AN
generations, respectively with $N=124$ and 367 nodes, and included
them into the groups consisting of networks with $N=100$ and 300. It
is important to recall that AN shows several features that are typical
both of small world (small $D$ and $\langle d\rangle$), and scale free
scenarios ($p(k)\sim k ^{-\gamma}$). The feature selection method can
help to identify whether AN lies closer to the WS or BA clusters and,
hence, to indicate which of the quoted scenarios it stays closer
to. The results $(v1,v2)=(-0.90,2.53)$ for $N=124$, and
$(-0.24,0.96)$ for $N=367$, indicate that, for both generations, the
AN is mapped away from all four clusters corresponding to each of the
considered theoretical models. This is, indeed, a very interesting
result, as it shows that our method is able to identify that this type
of network presents quite distinct topological features, albeit it
shares some common properties with those that fit into those the small
world and scale free scenarios.  In other words, the very fact that
several features of a given network coincide with those typical for a
large network class, does not automatically implies that it belongs to
the same set.

\section{Concluding Remarks}

Much of the advances in science have only been allowed by ability of
researchers to focus attention on the most important features and
variables in each problem.  Because human beings have a rather limited
ability to cope with a large numbers of measurements, it becomes
critical to devise and apply methods which can possibly identify the
most relevant features.  Fortunately, sound and objective concepts and
methods --- defining the research area called \emph{feature selection}
--- have been developed which can help us in such tasks.  Perhaps for
historical reasons, such methods are not so widely known and used by
the Physics community.

Because highly structured complex networks can only be comprehensively
characterized by considering several measurements, the application of
feature selection methods presents great potential for helping researchers
in that area.  In a recent work ~\cite{Costa_surv:2007}, canonical
variables projections and Bayesian decision theory were applied in order
to classify complex networks and to investigate measurements. The current
work has unfolded such a possibility with respect to the discriminative
potential of a comprehensive set of hierarchical measurements.

While the traditional node degree, clustering coefficient and shortest
path provide quantifications of important features of the networks
under analysis, they are degenerated in the sense that several
networks may map into the same measurement values.  The extension of
such concepts to reflect also the progressive neighborhoods around
each node has been proposed (e.g.~\cite{Costa_hier:2004,
Costa_Rocha:2005, Costa_Silva:2006, Costa_Sporns:2006, Andrade:2006})
in order to obtain enhanced, less degenerated, characterizations of
complex networks.  Each of such measurements are defined for a series
of hierarchical levels $\ell$, yielding a high dimensional measurement
space.  Actually, the values of such measurements taken at each level
can be understood as a measurement in itself. Given such a large
number of features, it becomes important to identify which
measurements are potentially more effective in providing
discriminative descriptions of specific types of networks under
analysis.  In this work, we applied standardization and canonical
projections in order to identify the most important measurements in
Table~\ref{tab:abbrs} with respect to four representative complex
networks models, namely Erd\"os-R\'enyi, Barab\'asi-Albert,
Watts-Strogatz and a geographical type of network.  Each hierarchical
measurement was calculated along three successive neighborhood levels.
The traditional average degree, clustering coefficient and shortest
path lengths were also considered.  A total of 13 measurements were
considered in our investigation.

Several interesting findings have been obtained by the applied
methodology.  First, four types of networks were well-separated even in
the two-dimensional canonical projected phase space (considerably worse
separations were obtained by PCA), with the pairs of models ER/BA and
WS/GG forming superclusters.  By taking into account the respective
weights of each measurement in the canonical projections, it was possible
to associate an overall importance value to each measurement.  Such values
were calculated for three network sizes ($N=100$, $200$ and $300$).  While
the traditional clustering coefficient was identified as contributing more
intensely for the separation between the network types, several
hierarchical measurements resulted in relatively high complementary
contributions, with the hierarchical clustering coefficient by balls
($C^B(\ell)$) and neighborhood clustering coefficient ($C^N(\ell)$)
providing particularly relevant contributions.  The node degree
(traditional and hierarchical), as well as the average shortest path
length, did not contribute significantly to the separation of the network
models.  It is important to recall that such results are, in principle,
specific to the separation of the four considered types of networks, in
the sense that different results may be obtained when considering other
networks models.

In addition to providing an objective means for selecting
measurements for characterization and discrimination of complex
networks models, the multivariate approach considered in this work
can also provide valuable insights about the structural differences
between distinct types of networks.  For instance, the fact that the
clustering coefficient resulted more relevant than the shortest path
length suggests that the four considered models present local
connectivity (expressed in the clustering coefficient) even more
distinct than shortest path length distribution.  It would be
interesting to apply such multivariate methods to the
characterization of other types of networks, especially those
involving community structure.

\vspace{1cm} {\bf Acknowledgment: } Luciano da F. Costa is grateful to
FAPESP (05/00587-5) and CNPq (301303/06-1) for financial support. Roberto
F. S. Andrade acknowledges support by FAPESB (012/05) and CNPq
(306369/04-4).

\bibliographystyle{apsrev}
\bibliography{mvr}

\begin{thebibliography}{21}
\expandafter\ifx\csname natexlab\endcsname\relax\def\natexlab#1{#1}\fi
\expandafter\ifx\csname bibnamefont\endcsname\relax
  \def\bibnamefont#1{#1}\fi
\expandafter\ifx\csname bibfnamefont\endcsname\relax
  \def\bibfnamefont#1{#1}\fi
\expandafter\ifx\csname citenamefont\endcsname\relax
  \def\citenamefont#1{#1}\fi
\expandafter\ifx\csname url\endcsname\relax
  \def\url#1{\texttt{#1}}\fi
\expandafter\ifx\csname urlprefix\endcsname\relax\def\urlprefix{URL }\fi
\providecommand{\bibinfo}[2]{#2}
\providecommand{\eprint}[2][]{\url{#2}}

\bibitem[{\citenamefont{Albert and Barab\'asi}(2002)}]{Albert_Barab:2002}
\bibinfo{author}{\bibfnamefont{R.}~\bibnamefont{Albert}} \bibnamefont{and}
  \bibinfo{author}{\bibfnamefont{A.~L.} \bibnamefont{Barab\'asi}},
  \bibinfo{journal}{Rev. Mod. Phys.} \textbf{\bibinfo{volume}{74}},
  \bibinfo{pages}{47} (\bibinfo{year}{2002}).

\bibitem[{\citenamefont{Dorogovtsev and Mendes}(2002)}]{Dorog_Mendes:2002}
\bibinfo{author}{\bibfnamefont{S.~N.} \bibnamefont{Dorogovtsev}}
  \bibnamefont{and} \bibinfo{author}{\bibfnamefont{J.~F.~F.}
  \bibnamefont{Mendes}}, \bibinfo{journal}{Advances in Physics}
  \textbf{\bibinfo{volume}{51}}, \bibinfo{pages}{1079} (\bibinfo{year}{2002}).

\bibitem[{\citenamefont{Newman}(2003)}]{Newman:2003}
\bibinfo{author}{\bibfnamefont{M.~E.~J.} \bibnamefont{Newman}},
  \bibinfo{journal}{SIAM Review} \textbf{\bibinfo{volume}{45}},
  \bibinfo{pages}{167} (\bibinfo{year}{2003}),
  \bibinfo{note}{cond-mat/0303516}.

\bibitem[{\citenamefont{Boccaletti et~al.}(2006)\citenamefont{Boccaletti,
  Latora, Moreno, Chavez, and Hwang}}]{Boccaletti_etal:2006}
\bibinfo{author}{\bibfnamefont{S.}~\bibnamefont{Boccaletti}},
  \bibinfo{author}{\bibfnamefont{V.}~\bibnamefont{Latora}},
  \bibinfo{author}{\bibfnamefont{Y.}~\bibnamefont{Moreno}},
  \bibinfo{author}{\bibfnamefont{M.}~\bibnamefont{Chavez}}, \bibnamefont{and}
  \bibinfo{author}{\bibfnamefont{D.-U.} \bibnamefont{Hwang}},
  \bibinfo{journal}{Physics Reports} \textbf{\bibinfo{volume}{424}},
  \bibinfo{pages}{175} (\bibinfo{year}{2006}).

\bibitem[{\citenamefont{da~F.~Costa et~al.}(2007)\citenamefont{da~F.~Costa,
  Rodrigues, Travieso, and Boas}}]{Costa_surv:2007}
\bibinfo{author}{\bibfnamefont{L.}~\bibnamefont{da~F.~Costa}},
  \bibinfo{author}{\bibfnamefont{F.~A.} \bibnamefont{Rodrigues}},
  \bibinfo{author}{\bibfnamefont{G.}~\bibnamefont{Travieso}}, \bibnamefont{and}
  \bibinfo{author}{\bibfnamefont{P.~V.} \bibnamefont{Boas}},
  \bibinfo{journal}{Adv. Phys.} \textbf{\bibinfo{volume}{56}},
  \bibinfo{pages}{167} (\bibinfo{year}{2007}).

\bibitem[{\citenamefont{da~F.~Costa et~al.}(2006)\citenamefont{da~F.~Costa,
  Kaiser, and Hilgetag}}]{Costa_out:2006}
\bibinfo{author}{\bibfnamefont{L.}~\bibnamefont{da~F.~Costa}},
  \bibinfo{author}{\bibfnamefont{M.}~\bibnamefont{Kaiser}}, \bibnamefont{and}
  \bibinfo{author}{\bibfnamefont{C.}~\bibnamefont{Hilgetag}}
  (\bibinfo{year}{2006}), \bibinfo{note}{physics/060727}.

\bibitem[{\citenamefont{Faloutsos et~al.}(1999)\citenamefont{Faloutsos,
  Faloutsos, and Faloutsos}}]{Faloutsos_etal:1999}
\bibinfo{author}{\bibfnamefont{M.}~\bibnamefont{Faloutsos}},
  \bibinfo{author}{\bibfnamefont{P.}~\bibnamefont{Faloutsos}},
  \bibnamefont{and}
  \bibinfo{author}{\bibfnamefont{C.}~\bibnamefont{Faloutsos}},
  \bibinfo{journal}{Comp. Comm. Rev.} \textbf{\bibinfo{volume}{29}},
  \bibinfo{pages}{251} (\bibinfo{year}{1999}).

\bibitem[{\citenamefont{Newman}(2001)}]{Newman:2001}
\bibinfo{author}{\bibfnamefont{M.~E.} \bibnamefont{Newman}},
  \bibinfo{journal}{cond-mat/0111070}  (\bibinfo{year}{2001}).

\bibitem[{\citenamefont{Kalisker et~al.}(2006)\citenamefont{Kalisker, Cohen,
  Mokryn, Dolev, Shavitt, and Havlin}}]{Cohen_etal:2006}
\bibinfo{author}{\bibfnamefont{T.}~\bibnamefont{Kalisker}},
  \bibinfo{author}{\bibfnamefont{R.}~\bibnamefont{Cohen}},
  \bibinfo{author}{\bibfnamefont{O.}~\bibnamefont{Mokryn}},
  \bibinfo{author}{\bibfnamefont{D.}~\bibnamefont{Dolev}},
  \bibinfo{author}{\bibfnamefont{Y.}~\bibnamefont{Shavitt}}, \bibnamefont{and}
  \bibinfo{author}{\bibfnamefont{S.}~\bibnamefont{Havlin}},
  \bibinfo{journal}{Phys. Rev. E} \textbf{\bibinfo{volume}{74}},
  \bibinfo{pages}{066108} (\bibinfo{year}{2006}),
  \bibinfo{note}{cond-mat/0305582}.

\bibitem[{\citenamefont{da~F.~Costa}(2004)}]{Costa_hier:2004}
\bibinfo{author}{\bibfnamefont{L.}~\bibnamefont{da~F.~Costa}},
  \bibinfo{journal}{Phys. Rev. Letts.} \textbf{\bibinfo{volume}{93}},
  \bibinfo{pages}{098702} (\bibinfo{year}{2004}).

\bibitem[{\citenamefont{da~F.~Costa and da~Rocha}(2005)}]{Costa_Rocha:2005}
\bibinfo{author}{\bibfnamefont{L.}~\bibnamefont{da~F.~Costa}} \bibnamefont{and}
  \bibinfo{author}{\bibfnamefont{L.~E.~C.} \bibnamefont{da~Rocha}},
  \bibinfo{journal}{Eur. Phys. J. B} \textbf{\bibinfo{volume}{50}},
  \bibinfo{pages}{237} (\bibinfo{year}{2005}).

\bibitem[{\citenamefont{da~F.~Costa and Silva}(2006)}]{Costa_Silva:2006}
\bibinfo{author}{\bibfnamefont{L.}~\bibnamefont{da~F.~Costa}} \bibnamefont{and}
  \bibinfo{author}{\bibfnamefont{F.~N.} \bibnamefont{Silva}},
  \bibinfo{journal}{J. Stat. Phys.} \textbf{\bibinfo{volume}{125}},
  \bibinfo{pages}{845} (\bibinfo{year}{2006}).

\bibitem[{\citenamefont{da~F.~Costa and Sporns}(2006)}]{Costa_Sporns:2006}
\bibinfo{author}{\bibfnamefont{L.}~\bibnamefont{da~F.~Costa}} \bibnamefont{and}
  \bibinfo{author}{\bibfnamefont{O.}~\bibnamefont{Sporns}},
  \bibinfo{journal}{Eur. Phys. J. B} \textbf{\bibinfo{volume}{48}},
  \bibinfo{pages}{567} (\bibinfo{year}{2006}),
  \bibinfo{note}{q-bio.NC/0508007}.

\bibitem[{\citenamefont{Andrade et~al.}(2006)\citenamefont{Andrade, Miranda,
  and Lob{\~{a}}o}}]{Andrade:2006}
\bibinfo{author}{\bibfnamefont{R.~F.~S.} \bibnamefont{Andrade}},
  \bibinfo{author}{\bibfnamefont{J.~G.~V.} \bibnamefont{Miranda}},
  \bibnamefont{and} \bibinfo{author}{\bibfnamefont{T.~P.}
  \bibnamefont{Lob{\~{a}}o}}, \bibinfo{journal}{Phys. Rev. E}
  \textbf{\bibinfo{volume}{73}}, \bibinfo{pages}{046101}
  (\bibinfo{year}{2006}).

\bibitem[{\citenamefont{Duda et~al.}(2001)\citenamefont{Duda, Hart, and
  Stork}}]{Duda_Hart:2001}
\bibinfo{author}{\bibfnamefont{R.~O.} \bibnamefont{Duda}},
  \bibinfo{author}{\bibfnamefont{P.~E.} \bibnamefont{Hart}}, \bibnamefont{and}
  \bibinfo{author}{\bibfnamefont{D.~G.} \bibnamefont{Stork}},
  \emph{\bibinfo{title}{Pattern Classification}}
  (\bibinfo{publisher}{Wiley-Interscience}, \bibinfo{year}{2001}).

\bibitem[{\citenamefont{McLachlan}(2004)}]{McLachlan:2004}
\bibinfo{author}{\bibfnamefont{G.~J.} \bibnamefont{McLachlan}},
  \emph{\bibinfo{title}{Discriminant Analysis and Statistical Pattern
  Recognition}} (\bibinfo{publisher}{Wiley}, \bibinfo{year}{2004}).

\bibitem[{\citenamefont{Han and Kamber}(2001)}]{Han_Kamber:2001}
\bibinfo{author}{\bibfnamefont{J.}~\bibnamefont{Han}} \bibnamefont{and}
  \bibinfo{author}{\bibfnamefont{M.}~\bibnamefont{Kamber}},
  \emph{\bibinfo{title}{Data Mining: Concepts and Techniques}}
  (\bibinfo{publisher}{Morgan Kaufmann}, \bibinfo{year}{2001}).

\bibitem[{\citenamefont{Hand et~al.}(2001)\citenamefont{Hand, Mannila, and
  Smyth}}]{Hand_etal:2001}
\bibinfo{author}{\bibfnamefont{D.}~\bibnamefont{Hand}},
  \bibinfo{author}{\bibfnamefont{H.}~\bibnamefont{Mannila}}, \bibnamefont{and}
  \bibinfo{author}{\bibfnamefont{P.}~\bibnamefont{Smyth}},
  \emph{\bibinfo{title}{Data Mining}} (\bibinfo{publisher}{The MIT Press},
  \bibinfo{year}{2001}).

\bibitem[{\citenamefont{da~F.~Costa and Jr.}(2001)}]{Costa_book:2001}
\bibinfo{author}{\bibfnamefont{L.}~\bibnamefont{da~F.~Costa}} \bibnamefont{and}
  \bibinfo{author}{\bibfnamefont{R.~M.~C.} \bibnamefont{Jr.}},
  \emph{\bibinfo{title}{Shape Analysis and Classification: Theory and
  Practice}} (\bibinfo{publisher}{CRC Press}, \bibinfo{year}{2001}).

\bibitem[{\citenamefont{Andrade et~al.}(2005)\citenamefont{Andrade, Herrmann,
  Andrade, and da~Silva}}]{Andrade:2005}
\bibinfo{author}{\bibfnamefont{J.~S.} \bibnamefont{Andrade}},
  \bibinfo{author}{\bibfnamefont{H.~J.} \bibnamefont{Herrmann}},
  \bibinfo{author}{\bibfnamefont{R.~F.~S.} \bibnamefont{Andrade}},
  \bibnamefont{and} \bibinfo{author}{\bibfnamefont{L.~R.}
  \bibnamefont{da~Silva}}, \bibinfo{journal}{Phys. Rev. Lett.}
  \textbf{\bibinfo{volume}{94}}, \bibinfo{pages}{018702}
  (\bibinfo{year}{2005}).

\bibitem[{\citenamefont{Doye and Massen}(2005)}]{Doye:2005}
\bibinfo{author}{\bibfnamefont{J.~P.~K.} \bibnamefont{Doye}} \bibnamefont{and}
  \bibinfo{author}{\bibfnamefont{C.~P.} \bibnamefont{Massen}},
  \bibinfo{journal}{Phys. Rev. E} \textbf{\bibinfo{volume}{71}},
  \bibinfo{pages}{016128} (\bibinfo{year}{2005}).

\end{thebibliography}
\end{document}